\documentclass[twocolumn]{aastex63}

\shorttitle{BH UCXBs: galactic low-frequency GW Sources}
\shortauthors{Qin, Jiang \& Chen}

\begin{document}


\title{Black Hole Ultra-compact X-ray Binaries: Galactic Low-Frequency Gravitational Wave Sources}


\author[0000-0001-9206-1641]{Ke Qin}
\affil{School of Physics, Zhengzhou University, Zhengzhou 450001, China}
\author[0000-0002-2479-1295]{Long Jiang}
  \affil{School of Science, Qingdao University of Technology, Qingdao 266525, China}
   \affil{School of Physics and Electrical Information, Shangqiu Normal University, Shangqiu 476000, China}
\author[0000-0002-0785-5349]{Wen-Cong Chen}
  \affil{School of Science, Qingdao University of Technology, Qingdao 266525, China}
  \affil{School of Physics and Electrical Information, Shangqiu Normal University, Shangqiu 476000, China}



\begin{abstract}
In the Galaxy, close binaries with compact objects are important low-frequency gravitational wave (GW) sources. As potential low-frequency GW sources, neutron star/white dwarf (WD) ultra-compact X-ray binaries (UCXBs) have been investigated extensively. Using the MESA code, we systematically explored the evolution of black hole (BH)-main sequence star (MS) binaries to diagnose whether their descendants can be detected by space-borne GW detectors. Our simulations show that BH-MS binaries with an initial orbital period less than the bifurcation period can evolve into BH UCXBs that can be detected by LISA. Such an evolutionary channel would form compact mass-transferring BH-WD systems rather than detached BH-WD systems. The calculated X-ray luminosities of BH UCXBs detected by LISA at a distance $d=1$ kpc are $\sim10^{33}-10^{35}~\rm erg\,s^{-1}$ ($\sim10^{34}-10^{35}~\rm erg\,s^{-1}$ for $d=10$ kpc), hence it is possible to detect their electromagnetic counterparts. It is worth emphasizing only some BH-MS systems with an initial orbital period very close to the bifurcation period can evolve toward low-frequency GW sources whose chirp masses can be measured. The maximum GW frequency of BH UCXBs forming by BH-MS pathway is about 3 mHz, which is smaller than the minimum GW frequency (6.4 mHz) of mass-transferring BH-WD originating from a dynamic process. Furthermore, we obtain an initial parameter space (donor-star masses and orbital periods) of progenitors of BH UCXB-GW sources, which can be applied to future population synthesis simulations. By a rough estimation, we predict that LISA could detect only a few BH UCXB-GW sources forming by the BH-MS channel.
\end{abstract}


\section{Introduction}
Since the Laser Interferometer Gravitational-Wave Observatory (LIGO) detectors detected the first high-frequency gravitational wave (GW) signal from GW150914 \citep{abbo16}, LIGO-Virgo detector network have already discovered decades double BHs merger events \citep{abbo21b}, and several double neutron stars (NSs) and BH-NS merger events \citep{abbo17,abbo21a}. These GW events provide enormous data in testing the gravitational theory, studying the nature of compact objects, and checking stellar and binary stars evolution theory.

The merger of double compact objects is a transient event, only providing a chance for the GW detection. However, the inspiral process of a close binary including compact object in the Galaxy could emit continuous low-frequency GW signals, which can be detected by a space-borne GW detector. Several space-borne GW detectors such as the Laser Interferometer Space Antenna (LISA) \citep{amar17,amar22}, TianQin \citep{luo16} and Taiji \citep{ruan20} are planning to be launched in the early 2030s.  Their scientific aims are to detect low-frequency GW signals with frequencies between $10^{-4}$ Hz and 0.1 Hz in the Galaxy, which originate from compact binaries with orbital periods less than 5 hours \citep{sluy11}. Up to now, several potential low-frequency GW sources in the Galaxy were extensively investigated in some compact binaries such as CVs, double white dwarfs \citep[WDs,][]{nele03,krem17,koro17,koro18,liu20}, AM CVn \citep{nele03,nele04,liu21,liu22}, NS-WD binaries \citep{taur18,chen21,yu21,pol21}, double NSs \citep{yu15,taur17,pol21}, NS ultra-compact X-ray binaries \citep[UCXBs,][]{chen20b,wang21,chen21}, and compact intermediate-mass BH X-ray binaries \citep{chen20a,han21}.

Due to an initial mass function, compact binaries including WDs are primary detection targets of space-borne GW detectors. For example, the expected detection number of double WDs as low-frequency GW sources is greater than $10^{5}$ \citep{nele01,ruit10,mars11,koro17}. Based on known numbers of binary radio millisecond pulsars, \cite{taur18} anticipated at least 100 NS-WD binaries as the potential Galactic LISA sources. In principle, the number of BH-WD binaries should be much smaller than that of NS-WD binaries. Therefore, little attention has so far been pursued to BH-WD binaries population. However, \cite{haaf12a} concluded that BH UCXBs (with typical orbital periods of 100-110 min) could also be formed by the mass transfer between a stellar-mass BH and a Roche-lobe overflowing helium WD. Population synthesis investigations predicted that the number of mass-transferring BH-WD binaries may exceed $10^{4}$ in the Galaxy \citep{hurl02,yung06}. \cite{shao21} estimated that $\sim1 $ BH-WD binaries could be detected in both electromagnetic and GW waveband based on a population synthesis simulation with a delayed model ( $\sim 20$ for a stochastic model).

BH-WD binary has still not been identified via electromagnetic radiation, however, they may be related to some important astrophysical phenomenons such as BH UCXBs \citep{haaf12a}, long gamma-ray bursts without supernova association \citep{dong18},
type I supernovae and subluminous type I supernovae \citep{wils04,metz12}, tidal disruption events \citep{kawa18,frag20}, and week(s)-long white transients \citep{bobr22}. Especially, mass-transferring BH-WD binaries were also proposed to potentially be detected by the LISA in mHz GW frequency band \citep{sber21}.

There exist several BH-WD binaries candidates. Transient AT2018kzr probably arised from a merger between an ONe WD and a NS or BH \citep{gill20}. The luminous X-ray source X-9 in globular cluster 47 Tucanae was also thought to be BH UCXB accreting from a WD \citep{mill15,bahr17,chur17,tudo18}. 
By evaluating the stability of mass transfer from a carbon-oxygen WD onto compact objects with different masses, \cite{chur17} concluded that the accretor of X9 in 47 Tuc must be a BH. BH-WD binaries mainly originated from dynamic processes channel and isolated binary evolution channel. In recent decades, a large number of binaries with compact objects had been found in globular clusters \citep{cami00,chom13,gies18}. These sources should formed through dynamic process channel such as direct collisions, tidal captures, and exchange interactions \citep{mack08,chat13,anto16,krem18b}. However, the dynamical processes can certainly be neglected in the Galactic field. For isolated binary evolution channel, the WD progenitors in BH-WD binaries may be main-sequence (MS) stars or He stars.

In the Galaxy, there exist ten BH low-mass X-ray binaries with a low-mass donor star ($\la 1~M_{\odot}$) and a short orbital period ($\la 0.5~\rm days$) \citep[see also Table 1 in][]{shao20}. These sources should evolved from BH-MS binaries with a low-mass or intermediate-mass donor star \citep{just06,chen06}. After the donor stars deplete their H-rich envelope due to a mass transfer, they would evolve into detached or semi-detached BH-WD binaries. In observation, the latter should appear as BH UCXBs and X-ray transients. Sixteen non-recurring X-ray binaries (12 quiescent, non-thermal X-ray sources and 4 single X-ray outburst transients) that have been identified near the Galactic center were argued to be BH low-mass X-ray binaries with an
orbital period range of $4-12~\rm hr$ \citep{hail18,mori21}. Due to selection effect, detached BH-WD binaries are difficult to be discovered because of the lack of bright outbursts for BH accretors. However, compact mass-transferring BH-WD binaries are potential low-frequency GW sources \citep{sber21}. Once a mass-transferring BH-WD binary is detected in low-frequency GW band, then we can catch more useful parameters by a targeted observation in electromagnetic band. Therefore, it is very significant to study the formation and evolution of BH-WD binaries in multi-messenger astrophysics era.

Shao \& Li (2021) performed detailed stellar evolution models for BH binaries with non-degenerate donor stars, and obtained parameter spaces for stable and unstable mass transfer under different metallicities. However, their work had not investigated the formation and evolution of BH UCXBs and low-frequency GW sources using a detailed stellar evolution simulation. Heretofore, a detailed modeling of the evolution of BH-MS binaries until the formation of BH UCXBs and low-frequency GW sources has been missing in the literature. In this work, we perform a detailed stellar evolution model for a large number of BH-MS binaries, and investigate whether or not their descendants will be visible as low-frequency GW sources. We first describe the stellar evolution code in Section 2. The simulated results are presented in Section 3. Finally, we give a brief discussion and conclusion in Sections 4 and 5, respectively.

\section{Stellar Evolution Code}
We utilize a binary update version in the Modules for Experiments in Stellar Astrophysics (MESA) code \citep[r12115, ][]{paxt11,paxt13,paxt15,paxt18,paxt19} to model the evolution of BH-MS binaries. At the beginning of simulation, a binary system containing a BH with a mass of $M_{\rm bh}=8~M_{\odot}$ and a MS companion star is assumed to exist in a circular and synchronized orbit. In calculations, the BH is regarded as a point mass, and the companion star is taken to be sun-like star with a mass range of $M_{\rm d}=0.4-2.0~ M_{\odot}$ and a solar composition (i.e. $X = 0.7, Y = 0.28, Z = 0.02$).
The MESA code continuously models the evolution of BH-MS binaries with various initial donor-star masses and initial orbital periods until the stellar age is greater than the Hubble time ($1.4\times10^{10}$ years) or the time step reaches a minimum time step limit.

In the MESA code, we use Type2 opacities for extra C/O burning during and after He burning. The mixing length is taken to be 2 times a local pressure scale height, i.e. $\alpha=2$. We adopt the "Ritter" mass-transfer scheme when the donor star fills its Roche lobe \citep{ritt88}. The default time step options with
$mesh_{-}delta_{-}coeff = 1.0$ and $varcontrol_{-}target = 10^{-4}$ are adopted. Our inlists are available at doi: \dataset[https://doi.org/10.5281/zenodo.7513654]{\doi{https://doi.org/10.5281/zenodo.7513654}}.

Orbital-angular-momentum loss from the binary system play a vital role in determining whether BH-MS binaries can evolve into low-frequency GW sources. In the simulation, we consider three types of orbital-angular-momentum loss mechanism producing by GW radiation,  magnetic braking, and mass loss. If the donor star develops a convective envelop (i.e. the donor-star mass is less than $1.5~ M_{\odot}$), the standard magnetic braking model with $\gamma=4$ is adopted \citep{verb81,rapp83}. The magnetic braking will stop working if the donor star loses its radiative core during the nuclear evolution. Once the donor star fills its Roche lobe, the mass transfer will be limited by the Eddington accretion rate
\begin{equation}
\dot{M}_{\rm Edd}=\frac{4\pi G M_{\rm bh}}{\kappa c \eta},
\end{equation}
where $G$ is the gravitational constant, $c$ is the speed of light in vacuo, $\kappa=0.2(1+X)$ ($X$ is the H abundance in the outer layers of the donor star) is the Thompson-scattering opacity of electrons.
The energy conversion efficiency of accreting BH is $\eta=1-\sqrt{1-(M_{\rm bh}/3M_{\rm bh,0})^{2}}$ \citep[for
$M_{\rm bh}<\sqrt{6}M_{\rm bh,0}$,][]{pods03}, here $M_{\rm bh,0}$ is the initial BH mass. The excess material above the Eddington accretion rate is thought to be ejected in the vicinity of the accreting BH, carrying away the specific orbital angular momentum of the BH.

When BH-MS binaries evolve into binary systems with a compact orbit, the emitting GW signals are possible to be caught by the space-borne GW detectors. Taking a mission duration of 4 years, we calculate the characteristic strain of BH binaries during inspirals of two components according to \citep{chen20a}
\begin{equation}
h_{\rm c}\approx 3.75\times 10^{-20}\left(\frac{f_{\rm gw}}{10^{-3}~\rm Hz}\right)^{7/6}\left(\frac{\mathcal{M}}{1~M_{\odot}}\right)^{5/3}\left(\frac{10~\rm kpc}{d}\right),
\end{equation}
where $f_{\rm gw}=2/P_{\rm orb}$ ($P_{\rm orb}$ is the orbital period) is the GW frequency, $d$ is the distance. For simplicity, the chirp mass $\mathcal{M}$ is estimated by
\begin{equation}
\mathcal{M}=\frac{(M_{\rm bh}M_{\rm d})^{3/5}}{(M_{\rm bh}+M_{\rm d})^{1/5}}.
\end{equation}
If our simulated characteristic strain exceed the sensitivity curve of LISA originating from an analytic estimation given by \cite{robs18}, the corresponding BH X-ray binaries are thought to be low-frequency GW sources.

In general, equation (3) can only calculate the chirp mass of a detached system, in which the orbital decay is fully caused by GW radiation. The mass transfer should contribute a reverse tendency of orbital evolution with GW radiation when the mass is transferred from the less massive donor star to the more massive BH. However, the mass transfer in BH binaries with a compact orbit is driven by GW radiation, and proceeds on a timescale near that of GW radiation \citep{haaf12a,haaf12b}. Therefore, equation (3) provides an approximate estimation for the chirp mass of compact BH binaries with a mass transfer.

\begin{figure}
\centering
\includegraphics[width=1.15\linewidth,trim={0 0 0 0},clip]{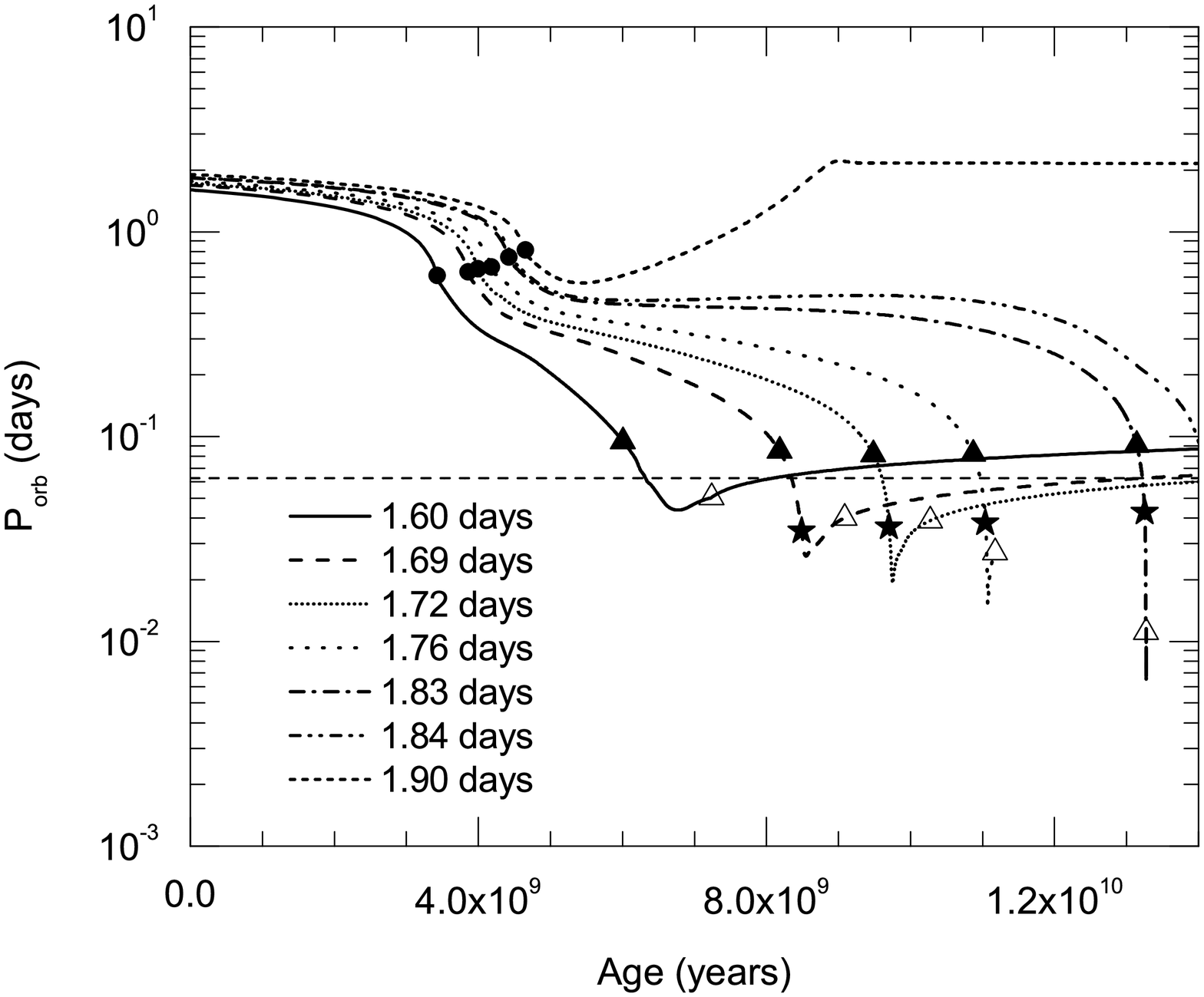}
\includegraphics[width=1.15\linewidth,trim={0 0 0 0},clip]{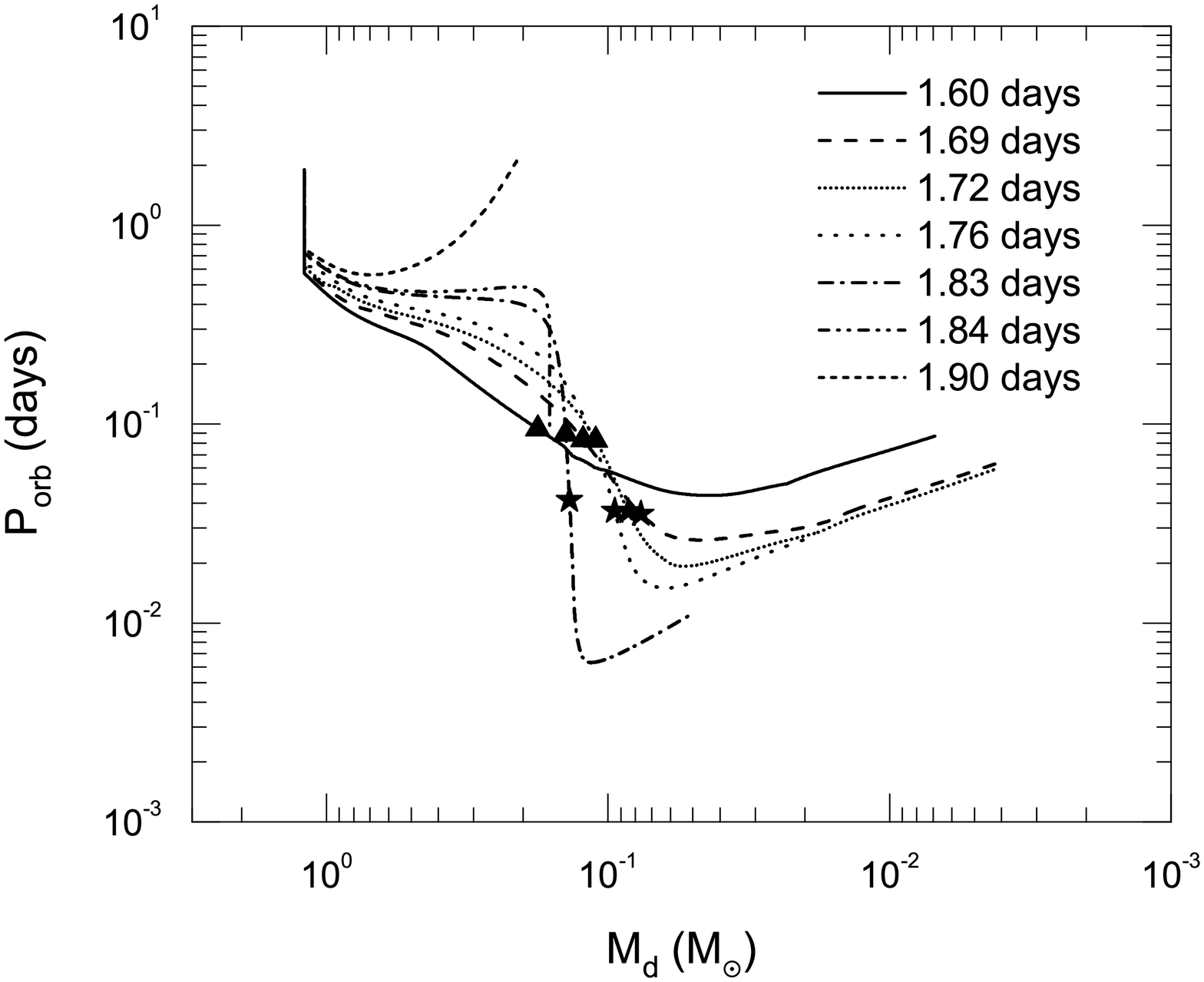}
\caption{Evolution of BH-MS binaries with an initial donor-star mass of 1.2 $M_{\odot}$ and different initial orbital periods in the orbital period vs. stellar age diagram (top panel), and the orbital period vs. donor-star mass diagram (bottom panel). The horizontal short dot line represents the threshold period \cite[90 minutes, see also][]{haaf13} of UCXBs. The solid circles denote the onset of mass transfer. The solid triangles and stars represent the onset that BH binaries could be detected by the LISA within a distance $d =1$, and 10 kpc, respectively. BH UCXBs will not be detected by LISA at a distance of 1 kpc after the open triangles. } \label{fig:orbmass}
\end{figure}

\section{Simulated Results}
To investigate the formation and evolution of BH-WD binaries, we first model the evolution of a BH-MS binary with an initial donor-mass of 1.2 $M_{\odot}$ for an initial orbital period of $P_{\rm orb, i}$= 1.60-1.90 days. Figure 1 displays the evolutionary tracks of seven BH-MS binaries in the orbital period versus stellar age plane, and the orbital period versus donor-star mass plane. Because the donor star possesses a convective envelope, the rapid orbital decay is caused by the magnetic braking in the early stage without a mass transfer. After the nuclear evolution of $\sim4$ Gyr (depending on the initial orbital period), the donor star fills its Roche Lobe, and triggers a mass transfer. The mass transfer tends to widen the orbit when the mass is transferred from the less massive donor star to the more massive BH, while the magnetic braking causes an opposite tendency. The competition between two mechanisms mentioned above determines the final fate of the binary: wide orbit system or narrow orbit system.

In Figure 1, the system with $P_{\rm orb, i}$= 1.90 days will evolve into a system with a wide orbit (the final orbital period is $\sim$ 2 days). When $P_{\rm orb, i}= 1.84-1.89$ days, the orbit of BH-MS binaries continuously shrink, while they can not become UCXBs within a Hubble time \citep[we take a threshold period of 90 minutes for UCXBs, see also][]{chen20b}. \cite{sluy05a} defined a bifurcation period as the
longest initial orbital period that results in compact binary systems with an ultra-short period within a Hubble time. According to this definition, the bifurcation period of BH-MS binaries with a 1.2 $M_{\odot}$ donor star is $P_{\rm bif}=1.83$ days. Actually, the bifurcation period is very sensitive to the angular-momentum-loss rate by magnetic braking (the influence of mass-loss mechanisms on bifurcation periods can be ignored). Comparing with the traditional magnetic braking, the saturated magnetic braking (which is a weak magnetic braking mechanism) would considerably decrease the bifurcation period \citep[see also,][]{sluy05b,ma09}. It is clear that BH-MS binaries with $P_{\rm orb, i}= 1.60-1.83$ days firstly appear as low-frequency GW sources that are visible by LISA at a distance of 1 kpc, then evolve into BH UCXBs with a maximum period of 90 minutes. It is worth noted that BH UCXBs can not always be detected by LISA at a distance of 1 kpc in entire UCXB stage, while they are certainly UCXBs if they can be detected by LISA at a distance of 10 kpc (see also solid triangles, solid stars, and open triangles in top panel of Figure 1). However, it requires a high GW frequency (corresponding to a short orbital period) in order to produce a similar characteristic strain for a similar chirp mass and long detection distance according to equation (2). To form such BH UCXBs with ultra-compact orbits, the donor stars must develop a more massive core by a long-term nuclear evolution, which corresponds to a relatively long initial orbital period. Therefore, a detection horizon of 10 kpc requires an relatively narrow range ($1.69-1.83$ days) of initial orbital periods. Same to \cite{chen20b}, an initial orbital period less than and near the bifurcation period would benefit to form more compact system, resulting in a far horizon of low-frequency GW detection.

For a detection horizon of 1 kpc, the orbital periods and the donor-star masses of BH UCXBs appearing as GW sources are approximately 0.1 days (which corresponds to a GW frequency of $\sim0.2$ mHz), and $\sim0.1-0.2~M_{\odot}$. It requires a short orbital period to produce strong chirp signals for a far detection horizon. To satisfy a long detection distance of 10 kpc, BH X-ray binaries would evolve into more compact systems with orbital periods of $\sim0.04$ days (which corresponds to a GW frequency of $\sim0.6$ mHz), and low donor-star masses of $0.08-0.14~M_{\odot}$. It is worth emphasizing that a short initial orbital period
(e.g. 1.60 days) is difficult to produce a low-frequency GW sources with a long detection horizon of 10 kpc.

\begin{figure}
\centering
\includegraphics[width=1.15\linewidth,trim={0 0 0 0},clip]{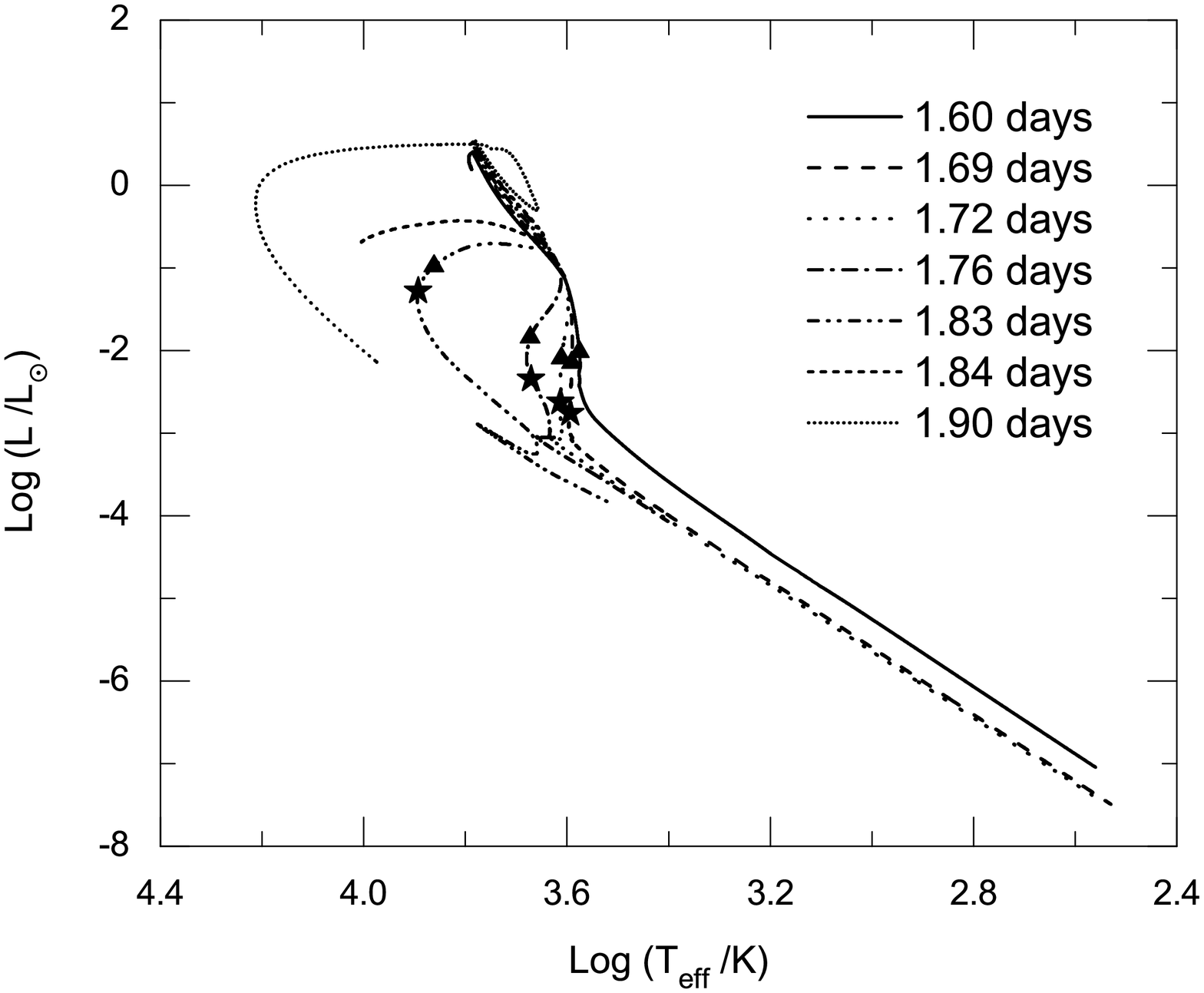}
\caption{Evolutionary tracks of BH-MS binaries with an initial donor star mass of 1.2 $M_{\odot}$ and different initial orbital periods in the H-R diagram. The solid triangles and stars represent the points that could be first detected by LISA within a distance $d =$ 1 kpc and 10 kpc, respectively.} \label{fig:orbmass}
\end{figure}

To understand the evolutionary properties of the donor stars, in Figure 1 we plot the evolutionary tracks of BH-MS binaries in H-R diagram. When BH X-ray binaries evolve into low-frequency GW sources, the luminosities of the donor stars are $\log(L/L_{\odot})\sim-1.0$, to -2.8, and the effective temperatures are in the range of 4000 to 8000 K. Such a luminosity range is consistent with that of WDs, however the effective temperatures are slightly lower than those of the observed WDs. Such a discrepancy should arise from the existence of thin H envelope on the surface of the donor stars (see also Table 1), which would lead to large radius and low effective temperature. It is worth emphasizing that a stable mass transfer can only produce semi-detached BH-WD systems rather than detached BH-WD systems. This is consistent with a conclusion that LISA might not detect any detached BH-WD systems in the Galaxy proposed by \cite{brei20}.

\begin{figure}
\centering
\includegraphics[width=1.15\linewidth,trim={0 0 0 0},clip]{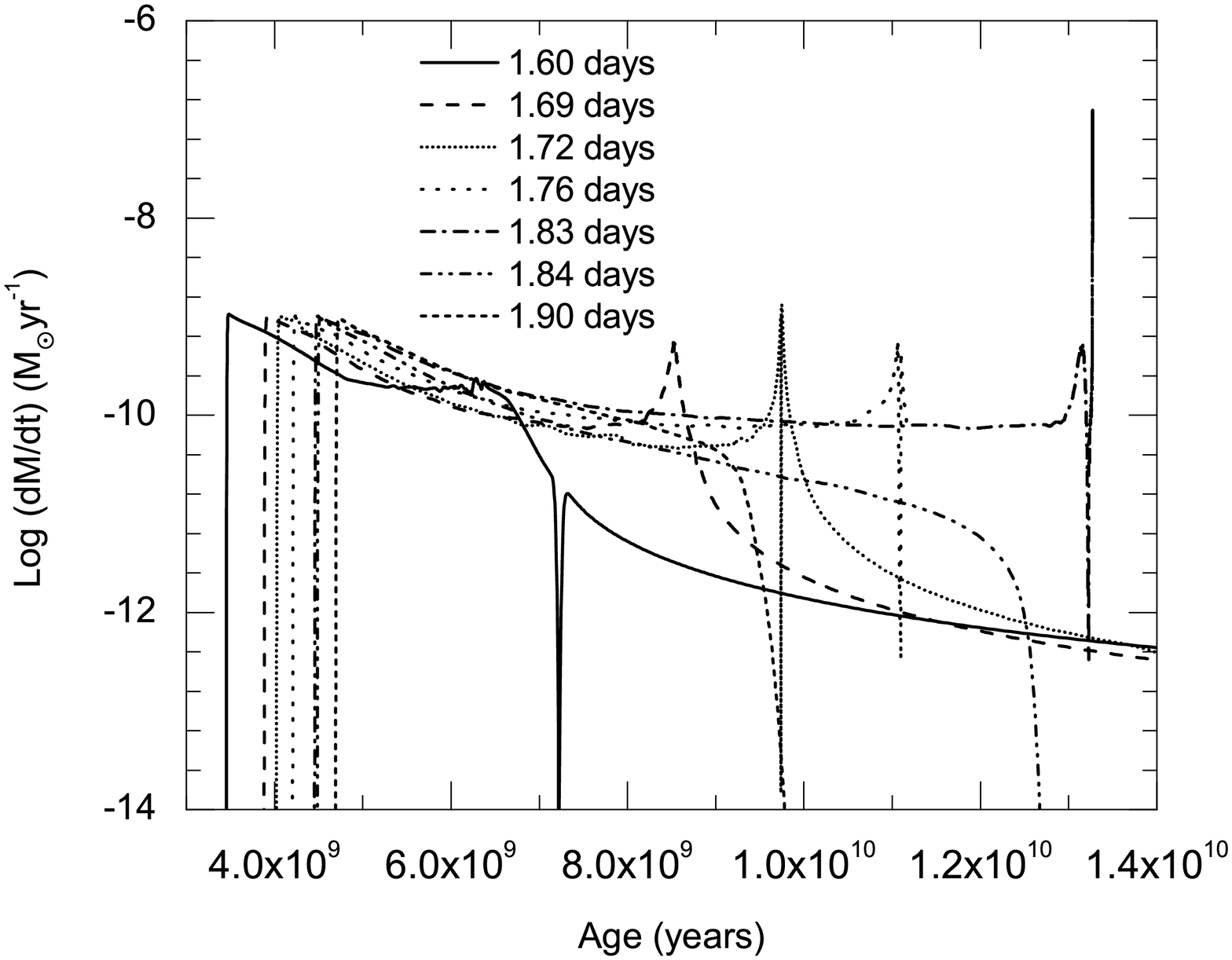}
\caption{Evolution tracks of the mass transfer rate of the donor star for BH-WD binaries in the mass transfer rate vs. stellar age diagram with an initial donor star mass of 1.2$M_{\odot}$ and different initial orbital periods.} \label{fig:orbmass}
\end{figure}

\begin{figure}
\centering
\includegraphics[width=1.15\linewidth,trim={0 0 0 0},clip]{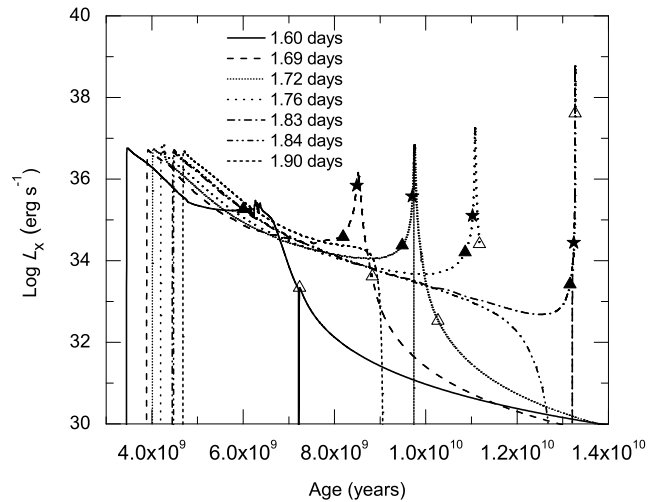}
\caption{Evolution of X-ray luminosity for BH X-ray binaries with an initial donor-star masses of 1.2 $M_{\odot}$. The solid triangles and solid stars represent the onsets that BH X-ray binaries can be detected by LISA within a distance $d =1$, and 10 kpc, respectively. BH UCXBs will not be detected by LISA at a distance of 1 kpc after the open triangles.} \label{fig:orbmass}
\end{figure}

In Figure 3, we plot the evolution of the mass transfer rate of BH X-ray binaries with an initial donor-star mass of 1.2$M_{\odot}$. After the nuclear evolution of $\sim4$ Gyr (depending on the initial orbital periods), the envelope of donor stars expand and fill their Roche lobes, and transfer the surface H-rich material onto the BHs. Mass transfer firstly proceed at a rate of $10^{-10}-10^{-9}~M_{\odot}\,\rm yr^{-1}$, which is obviously smaller than those ($10^{-8}-10^{-7}~M_{\odot}\,\rm yr^{-1}$) of BH X-ray binaries with intermediate-mass donor stars \citep{just06,chen06}. Once GW radiation dominates the orbital evolution of BH UCXBs, the orbits begin to rapidly shrink. When these sources evolve into the systems with minimum orbital period, the mass transfer rates emerge a peak, and then sharply decline. 

Figure 4 illustrates the evolution of X-ray luminosities of BH X-ray binaries. We calculate the X-ray luminosity of the accretion disk around the BH by expressions as follows \citep{kord02}:
\begin{equation}
L_{X}=\left\{
\begin{array}{cl}
\epsilon\dot{M}_{\rm acc}c^{2},\quad \dot{M}_{\rm crit}<\dot{M}_{\rm acc}\leq\dot{M}_{\rm Edd}\\
\epsilon\left(\frac{\dot{M}_{\rm acc}}{\dot{M}_{\rm crit}}\right)\dot{M}_{\rm acc}c^{2},\quad \dot{M}_{\rm acc}<\dot{M}_{\rm crit}\\
\end{array}\right.
\end{equation}
where $\epsilon=0.1$ is the radiative efficiency of the accretion disk around a BH, and the critical accretion rate is taken to be $\dot{M}_{\rm crit}=10^{-9}~\rm M_{\odot}\,yr^{-1}$ for a stellar-mass BH \citep{nara95,hopm04}.

Comparing with detached binaries such as double WD, double NS, and double BH, BH X-ray binaries are ideal multi-messenger objects in both low-frequency GW band and electromagnetic waves band. X-ray observations for these GW sources will provide many useful information and relatively accurate position. Before BH X-ray binaries evolve into low-frequency GW sources, their X-ray luminosities keep a relatively low state with a luminosities of $\sim10^{33-35}~\rm erg\,s^{-1}$. Once the systems appears as low-frequency GW source, the X-ray luminosities would rapidly climb due to a rapid shrinkage of the orbits. The  typical peak X-ray luminosity of BH X-ray binaries is about $10^{37-39}~\rm erg\,s^{-1}$ in transient bursts stages \citep{teta16}. When $P_{\rm orb,i}=P_{\rm bif}$, the X-ray luminosity of the BH X-ray binary is relatively low when it begin to appear as low-frequency GW sources. Subsequently, it would rapidly evolve toward luminous sources with a maximum X-ray luminosity of $6.31\times10^{38}~\rm erg\,s^{-1}$, which is close to that of the so-called the ultra-luminous X-ray sources (ULXs; $L_{X}\geq10^{39}~\rm erg\,s^{-1}$) \citep{feng11,kaar17}. It should noted that such a high X-ray luminosity requires a fine-tuning $P_{\rm orb,i}$ ($=P_{\rm bif}$) and its high X-ray luminosity is transient. The X-ray flux detection limit of non-focussing X-ray telescopes is $\sim10^{-11}~\rm erg\,s^{-1}\,cm^{-2}$ \citep{sen21}, which yields a minimum detectable luminosity of $\sim10^{33}~\rm erg\,s^{-1}$ and $\sim10^{35}~\rm erg\,s^{-1}$ for a distance of 1 kpc and 10 kpc \citep{vanb20}, respectively. According to our simulations, BH X-ray binaries are mostly likely to be observable in both low-frequency GW band and X-ray band (besides early LISA source stage of BH X-ray binary with $P_{\rm orb,i}=1.83$ days at a distance of 10 kpc and later LISA source stage of BH X-ray binary with $P_{\rm orb,i}=1.72$ days at a distance of 1 kpc).

Table 1 lists some important evolutionary parameters of BH-MS binaries that can evolve into low-frequency GW sources. At the moment of Roche lobe overflow, the donor stars in systems with a long initial orbital period have a low center H abundance, which is caused by a long timescale nuclear evolution. If our simulated characteristic strains exceed an analytic estimation for the sensitivity curve of LISA given by \cite{robs18}, the corresponding BH X-ray binaries are thought to be detectable as low-frequency GW sources. The detection timescales (between the intersections of the different evolutionary tracks with the horizontal line and open triangles in top panel of Figure 1) that BH UCXBs can be detected by LISA within a distance of 1 kpc are in the range from 53.6 to 916 Myr (it is worth noted that these timescales depend on the threshold period of UCXBs), while these timescales are 21.4 $-$ 95.1 Myr for a detection horizon of 10 kpc. BH-MS binaries with a long initial orbital period tend to appear as low-frequency GW sources with a short detection timescale and a low X-ray luminosity.

\begin{figure}
\centering
\includegraphics[width=1.15\linewidth,trim={0 0 0 0},clip]{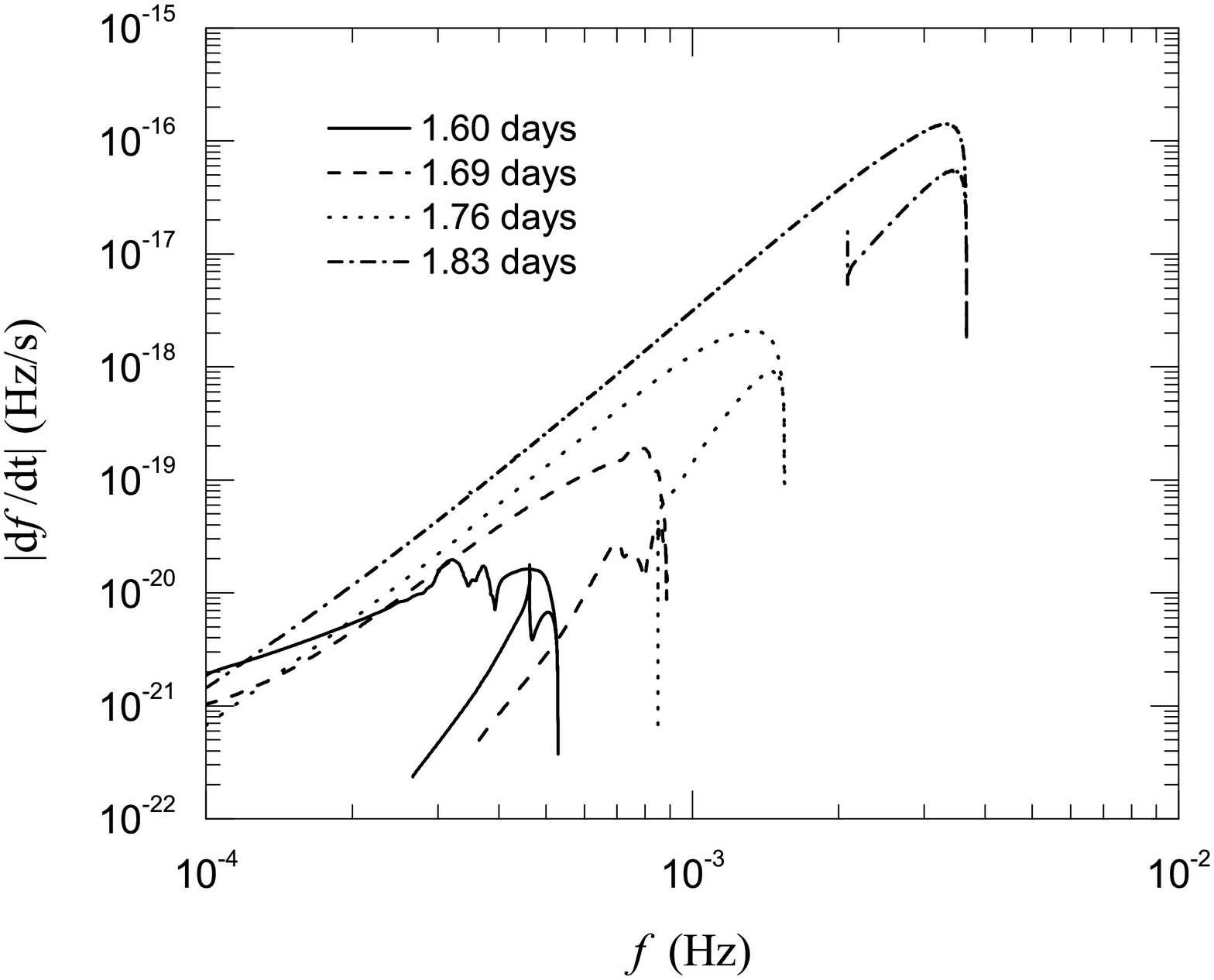}
\caption{Evolution of the frequency derivative of low-frequency GW signals emitting by BH UCXBs with an initial donor star mass of 1.2 $M_{\odot}$ and different initial orbital periods in the $|\dot{f}|$ vs. $f$ diagram. } \label{fig:orbmass}
\end{figure}

In principle, all BH X-ray binaries always emit GW signals due to a change of their mass quadrupole moment in the inspiral process. However, it strongly depends on the GW frequency and characteristic strain whether these source can be detected by LISA. Figure 5 presents the evolution of GW frequency derivative ($\dot{f}$) as a function of GW frequency. With the rapidly shrinkage of the orbit, angular-momentum-loss rate continuously increases, resulting in a rapid climb of GW frequency derivative. All curves in the climbing stage are approximately lines. The line scope of 1.83 days can be estimated to be $n=\Delta{\rm log}\dot{f}/\Delta{\rm log}f\approx 3.4$. According to equation (6), $n_{0}={\rm d\, log}\dot{f}/{\rm d\, log}f= 11/3$ for detached binary systems whose orbital decay is fully driven by GW radiation. Since $n$ is very close to $n_{0}$, it implies that the GW radiation is the dominant mechanism driving the orbital evolution of BH UCXBs with an initial orbital period very close to the bifurcation period.

\begin{table*}
\begin{center}
\caption{Some important evolutionary parameters for BH-MS binaries with an initial donor-star mass of $1.2~M_{\odot}$ and different initial orbital periods. \label{tbl-2}}
\begin{tabular}{@{}lllllllllllll@{}}
\hline\hline\noalign{\smallskip}
$P_{\rm orb,i}$ & $t_{\rm rlov}$ &  $X_{\rm rlov}$ &$M_{\rm d,1}$ &$P_{\rm orb,1}$ & $M_{\rm H,1}$ &$L_{\rm X, 1 }$&$M_{\rm d,10 }$ &$P_{\rm orb,10}$ & $M_{\rm H,10}$ &$L_{\rm X, 10}$&$\bigtriangleup t_{\rm LISA,1}$&$\bigtriangleup t_{\rm LISA,10}$ \\
(days) & (Gyr) &   & ($ M_{\odot}$)     & (days) &($ M_{\odot}$)  & ($\rm erg\,s^{-1}$)   & ($ M_{\odot}$)     & (days) &($ M_{\odot}$)  & ($\rm erg\,s^{-1}$) &(Myr)  & (Myr)\\
\hline\noalign{\smallskip}

1.83& 4.41 & $5.5\times10^{-5}$ & 0.141  & 0.0884 & 0.0231 & $2.4\times10^{33}$ & 0.137 &0.0426 & 0.0188 &$5.2\times10^{34}$& 53.6&21.4\\

1.76& 4.17 & 0.0266             & 0.113  & 0.0827  & 0.0593 & $1.7\times10^{34}$ & 0.0953 &0.0377 & 0.0419 &$3.1\times10^{35}$& 207&80.6\\

1.72& 3.99 & 0.0639             & 0.110  & 0.0824  & 0.0759 & $2.6\times10^{34}$ & 0.0847 &0.0363 & 0.0503 &$2.7\times10^{35}$& 660&94.7\\

1.69& 3.86 & 0.0936             & 0.124  & 0.0849  & 0.104 & $4.9\times10^{34}$ & 0.0757 &0.0349 & 0.0565 &$6.7\times10^{35}$& 736&95.1\\

1.60& 3.44 & 0.195             & 0.177  & 0.0943  & 0.177 & $2.0\times10^{35}$ & $-$ &$-$ & $-$ &$-$& 916&$-$\\

\hline\noalign{\smallskip}
\end{tabular}
\tablenotetext{}\\{Note. The columns list (in order): the initial orbital period, the stellar age and the hydrogen abundance at the beginning of Roche lobe overflow, the donor-star mass, orbital period, hydrogen envelope mass and X-ray luminosity when BH X-ray binaries are detected by LISA at a distance of 1 kpc, the donor-star mass, orbital period, hydrogen envelope mass and X-ray luminosity when BH UCXBs are detected by LISA at a distance of 10 kpc, the detection timescale that BH UCXBs can be detected by LISA within a distance of 1 kpc, and 10 kpc, respectively.}
\end{center}
\end{table*}

In Figure 5, the maximum $\dot{f}$ is close to the maximum GW frequency, while it is not exactly consistent with the minimum orbital period. When the GW frequency $f$ reaches a maximum, $\dot{f}$ should sharply declines (in principle, $\dot{f}=0$ at the maximum $f$), and then its sign changes into negative due to an orbital expansion. The maximum $\dot{f}$ for BH UCXBs with an initial orbital period of 1.83, 1.76, 1.69, and 1.60 days approximately are $10^{-16}$, $3\times10^{-18}$,  $10^{-19}$, and $10^{-20}~\rm Hz\,s^{-1}$, respectively. A detailed estimation for the detection limit on the minimum $\dot{f}$ is given by \citep{taka02,taur18}
\begin{equation}
\dot{f}_{\rm min}\sim2.5\times10^{-17}\left(\frac{10}{\rm SNR}\right)\left(\frac{4 \rm yr}{T}\right)^{2}~\rm Hz\,s^{-1},
\end{equation}
where SNR is the signal-to-noise ratio of GW signals. Meanwhile, the measurement of the chirp mass must employ a detection of $\dot{f}$ by an expression as
\begin{equation}
\mathcal{M}=\frac{c^{3}}{G}\left(\frac{5\pi^{-8/3}}{96}f^{-11/3}\dot{f}\right)^{3/5}.
\end{equation}
Therefore, it seems that only $\dot{f}$ of BH X-ray binaries with an initial orbital period very near the bifurcation period can be detected, and the chirp mass of corresponding systems can be derived. Actually, it is expected that $\dot{f}$ and chirp mass can be measured in about 25\% of the resolvable LISA sources \citep{amar12}.

\begin{figure}
\centering
\includegraphics[width=1.15\linewidth,trim={0 0 0 0},clip]{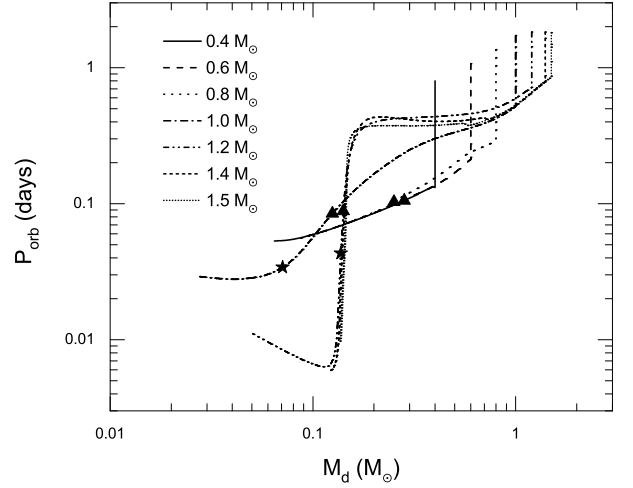}
\caption{Evolution tracks of BH-WD binaries with different initial donor-star masses and $P_{\rm orb,i}=P_{\rm bif}$ in the $P_{\rm orb}-M_{\rm d}$ diagram. The solid triangles and stars respectively represent the points that could be first detected by LISA within a distance d = 1 kpc and 10 kpc.} \label{fig:orbmass}
\end{figure}

The evolution of BH-MS binaries with different initial donor-star masses and an initial orbital period equaling to the bifurcation period in the orbital period vs. donor-star mass plane in Figure 6. The solid triangles and solid stars represent the onset stage that BH X-ray binaries can be detected by LISA within a distance of 1 kpc and 10 kpc. A massive donor star with a mass of 1.0-1.5 $M_{\odot}$ tends to evolve into a more compact system, which has a long detection horizon of 10 kpc for LISA. However, BH X-ray binaries with an initial donor-star mass of 0.4-0.8 $M_{\odot}$ can only evolve into GW sources with a detection horizon of 1 kpc, which have minimum orbital period of 0.06 days. Because donor stars with a small initial mass have a long evolutionary timescale  before they fill their Roche lobe, it results in the formation of a massive compact He core. Therefore, donor stars with a small initial mass (0.4-0.8 $M_{\odot}$) tend to possess a high donor-star mass ($\sim0.2-0.3 ~M_{\odot}$) when BH UCXBs are visible by LISA in a distance of 1 kpc. For an initial donor star mass of 1.2 $M_{\odot}$, the orbital periods are approximately 0.1 day and 0.04 day when BH UCXBs can be detected by LISA at a distances of 1 kpc and 10 kpc, and both cases have donor stars of about 0.1 $M_{\odot}$. For binaries with different initial donor-star masses and $P_{\rm orb,i}=P_{\rm bif}$, the orbital periods are approximately 0.09 day and 0.04 day when they evolve into LISA source at a distance $d = 1$ kpc and 10 kpc, respectively.

\begin{figure*}
\centering
\includegraphics[width=2.0\columnwidth]{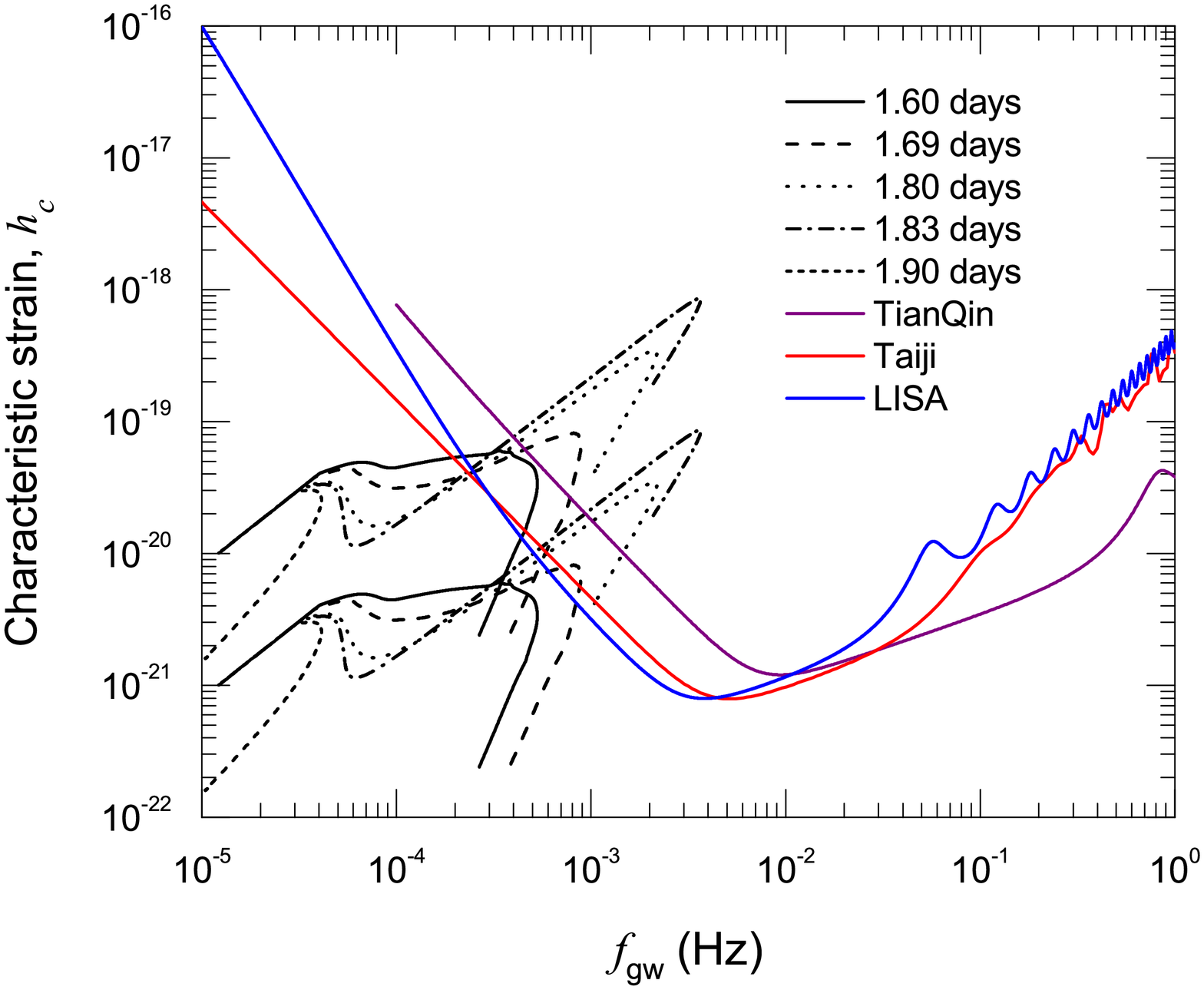}
\caption{Evolution of BH-MS binaries with an initial donor-star mass of 1.2 $M_{\odot}$ and different initial orbital periods in the characteristic strain vs. GW frequency diagram. The upper and under curve groups correspond to a detection distance of 1 kpc and 10 kpc, respectively. The blue, red, and purple curves represent the sensitivity curve of LISA, Taiji \citep{ruan20}, and TianQin \citep{wang19} arising from a numerical calculation 4 yr of observations, respectively.} \label{fig:orbmass}
\end{figure*}

\begin{figure*}
\centering
\includegraphics[width=2.0\columnwidth]{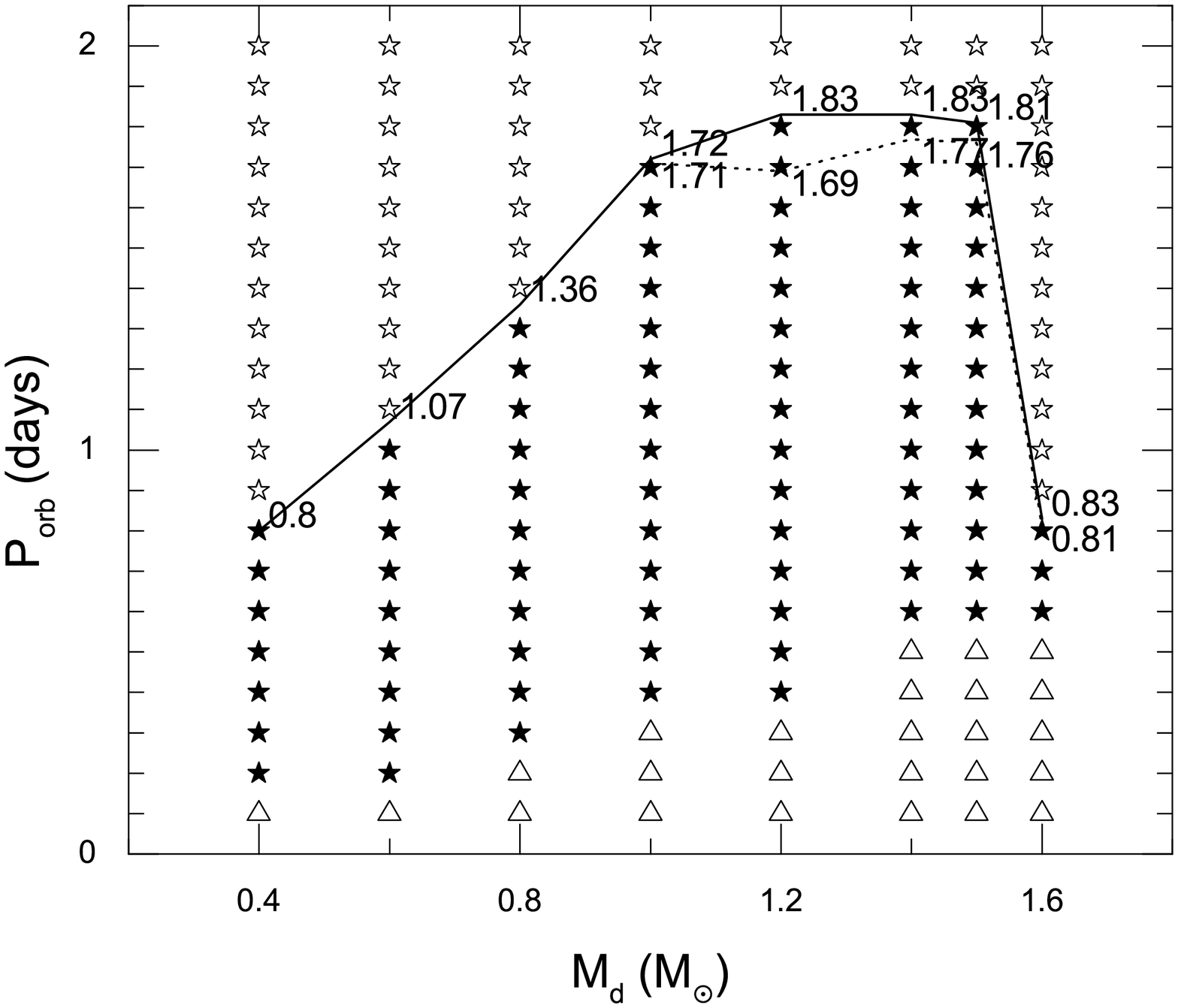}
\caption{Parameter space distribution of BH-MS binaries with different evolutionary fates in the initial orbital period vs. initial donor-star mass diagram. The initial masses of the BHs are taken to be $8 ~ M_{\odot}$. The solid curve denotes the bifurcation periods of BH-MS binaries with different donor-star masses, while the dashed curve represents the minimum initial orbital period of the progenitors of BH UCXBs that could be detected by LISA within a distance $d = 10$ kpc. Numbers inside the curves represent the initial orbital periods in units of days. The open stars implies that the corresponding BH-MS binaries will evolve toward systems with wide orbits. The solid stars correspond to BH-MS binaries that will evolve toward low-frequency GW sources detecting by LISA within a distance $d = 1$ kpc. The open triangles denote the binaries that the donor stars have already filled their Roche lobes at the beginning of binary evolution.} \label{fig:distribution}
\end{figure*}

Figure 7 shows the evolutionary tacks of BH X-ray binaries with an initial donor-star mass of 1.2 $M_{\odot}$ and different initial orbital period in the characteristic strain vs. GW frequency diagram. One can see that our simulated four BH UCXBs with an initial orbital period less than the bifurcation period can be detected by LISA and Taiji at a distance of 1kpc. If the detection distances prolong to be 10 kpc, only those systems with $P_{\rm i}=1.69-1.83$ days are visible by LISA and Taiji. Because of a short arm length, the detectability of TianQin on BH UCXBs is slightly weaker than that of LISA and Taiji. When the detection distance is $d=1~\rm kpc$, BH UCXB with an initial orbital period equaling to the bifurcation period produce a maximum characteristic stain of $\sim10^{-18}$ at a maximum GW frequency of $f\approx3~\rm mHz$. Such a maximum characteristic stain is similar to that of neutron star UCXBs, while the maximum GW frequency is slightly smaller than that ($f\approx4-5~\rm mHz$) of neutron star UCXBs \citep{taur18,chen20b}. Such a difference should originate from a big Roche-lobe radius of BHs, resulting in a wide orbital separation.

To understand the progenitor properties of BH X-ray binaries that can evolve into low-frequency GW sources detecting by LISA, we have simulated the evolution of a large number of BH-MS binaries in order to obtain the initial distribution of donor-star masses and orbital periods. The final evolutionary fates of our simulated BH-MS binaries are summarized in Figure 8. The solid curve represents the bifurcation period, over which the descendants of BH-MS binaries would evolve into X-ray binaries with wide orbits, or evolve toward compact binaries that can not be detected by LISA in a Hubble time. Except for some BH-MS binaries with donor stars filled the Roche lobe at the beginning of binary evolution, other binaries with an initial orbital period less than the bifurcation period can evolve into low-frequency GW source with a detection horizon of 1 kpc. Since BH-MS binaries with an initial orbital period near the bifurcation period tend to evolve into ultra-compact systems, LISA GW sources with a detection horizon of 10 kpc have an extremely narrow parameter space, which is enclosed by solid curve and dashed curve. It is clear that the initial donor-star mass less than 1.0 $M_{\odot}$ is impossible to result in the formation of LISA GW sources detecting at a distance of 10 kpc due to a long evolutionary timescale before the donor star filled its Roche lobe. It is worth noting that BH-MS binaries with a 1.6 $M_{\odot}$ donor star can also evolve into LISA GW sources. According to the standard magnetic braking model, donor stars with a mass greater than  1.5 $M_{\odot}$ are not expected to experience magnetic braking \citep{verb81,rapp83}, and firstly evolve toward long orbital period. However, once the donor stars develop a convective envelope the magnetic braking can still operate, then drive the binaries to evolve into short period population.
\section{Discussion}
\subsection{isolated binary evolution channel and dynamic process channel}
Our simulations show that a stable mass transfer in BH-MS binaries can not form compact detached BH-WD systems. However, BHs might capture WDs through dynamic process, then assemble compact BH-WD binaries. Due to the GW radiation, their orbits continuously shrink, and detached BH-WD binaries formed from the dynamic process will be detected by space-borne GW detectors. According to our simulation, such GW sources should arise from the dynamic process channel if their electromagnetic counterpart can not be identified (it is worth emphasizing that several semi-detached BH-WD binaries are difficult to detect X-ray emission in their entire GW source stage because X-ray flux is below the detection threshold, see also Figure 4). However, BHs would disrupt WDs once they enter their tidal radius \citep{hopm04}
\begin{equation}
R_{\rm t}=\left(\frac{M_{\rm bh}}{M_{\rm wd}}\right)^{1/3}R_{\rm wd},
\end{equation}
where $M_{\rm wd}$, and $R_{\rm wd}$ are the mass and radius of WDs, respectively. PTF J0533+0209 was reported to be a double WD
system with a short orbital period of 20 minutes as LISA verification source. Its He WD mass and radius are $\sim0.2~M_{\odot}$ and $\sim0.06~R_{\odot}$ \citep{burd19}. Taking $M_{\rm wd}=0.2~M_{\odot}$, $M_{\rm bh}=8~M_{\odot}$, and $R_{\rm wd}=0.06~R_{\odot}$, the tidal radius is estimate to be $R_{\rm t}=0.2~R_{\odot}$. When BHs accrete from the disrupted WDs, the systems appears as BH UCXBs. Assuming that the tidal radius is equaling to the orbital separation $a$, the minimum GW frequency from BH UCXBs formed by the dynamic process channel is
\begin{equation}
f_{\rm min}=\frac{2}{P_{\rm orb}}=\left[\frac{G(M_{\rm bh}+M_{\rm wd})}{\pi^{2}a^{3}}\right]^{1/2}\approx6.4~\rm mHz.
\end{equation}
Such a GW frequency obviously exceed the maximum frequency ($\approx3~\rm mHz$) emitting from BH UCXBs evolving from BH-MS binaries by a stable mass transfer. Therefore, the GW frequency can be used to confirm the formation channel of BH UCXBs.

\subsection{effective temperatures of mass-transferring WDs}
According to our models, it is impossible to form detached BH-WD binaries from BH binaries with a low-mass donor star via isolated binary evolution channel. Our simulated effective temperatures of mass-transferring WDs are in the range of 4000-8000 K when BH X-ray binaries appear as low-frequency GW sources detecting by LISA. These temperatures are slightly smaller than the observed lower limit of 8000 K. Our simulated donor stars in BH UCXB-GW sources possess hydrogen envelope with a mass of $0.02-0.17~M_{\odot}$. The existence of these hydrogen envelope would lead to a large radius and a low effective temperature. It is noteworthy that similar discrepancy was also found in simulating the formation of PTF J0533+0209. \cite{chen22} can reproduce the observed WD masses, radius of He WD, and orbital period of PTF J0533+0209, while their calculated effective temperature (6689 K) is obviously lower than the observed value ($20000\pm800~\rm K$). Strong H-shell flashes during the evolution of the He WD was anticipated to be most plausible mechanism resulting in such a WD with a high effective temperature \citep{chen22}.


\subsection{anomalous magnetic braking}
Because intermediate-mass stars without convective envelope are not generally expected to experience magnetic braking, in this work we only consider BH UCXBs evolving from BH-MS binaries with a low-mass donor star. However, a small fraction of intermediate-mass stars (so-called Ap and Bp stars) possess anomalously strong surface magnetic fields \citep[$10^{2}-10^{4}~\rm G,$][]{moss89,brai04}. By a magnetic couple between irradiation-driven stellar winds and strong surface magnetic fields, these Ap/Bp stars were thought to undergo an anomalous magnetic braking \citep{just06}. \cite{chen16} found that the anomalous magnetic braking of Ap/Bp stars can interpret three of persistent NS UCXBs with a long orbital-period and a high mass-transfer rate. However, only a small fraction ($\sim5\%$) of A/B stars have anomalously strong magnetic fields \citep{land82,shor02}. Furthermore, the number of intermediate-mass stars should be obviously smaller than that of low-mass stars according to an initial mass function (IMF). Therefore, the contribution of BH-MS binaries with an intermediate-mass donor star on BH UCXBs-GW sources can be ignored.

\subsection{detectability of BH UCXB-GW sources}
By a binary population synthesis simulation, \cite{pods03} obtained a formation rate of BH binaries with a donor star in the mass range of $0.5-2~M_{\odot}$ to be $\sim6\times10^{-7}~\rm yr^{-1}$ when the envelope structure parameter is $\lambda=0.5$ (see also their figure 3). The fraction of our orbital-period range ($d=1~\rm kpc$) overlapping their range for donor stars less than $2~M_{\odot}$ is about 90\% \citep[see also Figure 1a of][]{pods03}. Taking the radius and the scale height of the Galaxy are 15 and 1 kpc, and  a uniform distribution of BH-MS binaries with a donor star less than $2~M_{\odot}$ in the Galactic disk, their birthrate within a distance of 1 kpc is estimated to be $\sim6\times10^{-7}~\rm yr^{-1}\times\frac{1^{2}}{15^{2}}=2.7\times10^{-9}~\rm yr^{-1}$. Our simulated donor-star masses of the progenitors of BH UCXB-GW sources are in the range of $0.4-1.6~M_{\odot}$. In the following estimation, we take a mass range of $0.5-1.6~M_{\odot}$ in order to use the results of \cite{pods03}. Similar to \cite{sen22}, we also consider an weight factor originating from an IMF. Using the IMF given by \citep{krou93,krou03}, the weight factor of low-mass stars with $0.5-1.0~M_{\odot}$ and $1.0-1.6~M_{\odot}$ are $W_{0.5-1.0}=\int^{1.0}_{0.5}m^{-2.2}{\rm d}m/(\int^{1.0}_{0.5}m^{-2.2}{\rm d}m+\int^{1.6}_{1.0}m^{-2.7}{\rm d}m)\approx0.8$ and $W_{1.0-1.6}=1-W_{0.5-1.0}\approx0.2$, respectively. Assuming initial orbital periods for BH-MS binaries with a low-mass donor star and a short orbital period obey a uniform distribution, the birth rate of BH UCXB-GW sources within a distance of 1 kpc can be estimated to be $R_{1}\sim0.9\times(W_{0.5-1.0}+\frac{1.6-1.0}{2.0-1.0}W_{1.0-1.6})\times(2.7\times10^{-9})\approx2.2\times10^{-9}~\rm yr^{-1}$ according to our simulated parameter space (see also figure \ref{fig:distribution}). Adopting a mean detection timescale $\bigtriangleup t_{\rm LISA,1}=515~\rm Myr$ (see also Table 1), the detection number of BH UCXB-GW sources within a distance of 1 kpc is $N_{1}=\bigtriangleup t_{\rm LISA,1}R_{1}\approx 1$.

According to figure \ref{fig:distribution}, there exist about five systems that can evolve into low-frequency GW sources detecting by LISA at a distance of 10 kpc, while this number is 85 for a detection distance of 1 kpc. Therefore, the birth rate of BH UCXB-GW sources in a distance range from 9 to 10 kpc is approximately $R_{9,10}\sim0.9\times(W_{0.5-1.0}+\frac{1.6-1.0}{2.0-1.0}W_{1.0-1.6})\times\frac{10^{2}-9^{2}}{15^{2}}\times\frac{5}{85}\times(6.0\times10^{-7})\approx2.6\times10^{-9}~\rm yr^{-1}$. Similar to \cite{chen20b}, $R_{i-1,i}\sim2.6\times10^{-9}~\rm yr^{-1}$ if the birthrate of BH UCXB-GW sources per kpc interval from $i-1$ kpc to $i$ kpc obey a constant distribution. The mean detection timescale $\bigtriangleup t_{\rm LISA,10}\approx70~\rm Myr$ for BH UCXB-GW sources in a distance range from 9 to 10 kpc, i.e. $\bigtriangleup t_{\rm LISA,10}\approx\bigtriangleup t_{\rm LISA,1}/10$. Assuming that the mean detection timescale of GW sources in a distance range from $i-1$ kpc to $i$ kpc is $\bigtriangleup t_{\rm LISA,i}\approx\bigtriangleup t_{\rm LISA,1}/i$, the detection number of BH UCXB-GW sources forming by the isolated binary evolution channel in the Galaxy can be roughly estimated to be $N=\sum_{i=1}^{15}R_{i-1,i}\bigtriangleup t_{\rm LISA,1}/i\approx4$.
Such an estimation is consistent with the prediction by a population synthesis simulation with a delayed model, in which $\sim1 $ BH-WD binaries could be detected in both electromagnetic and GW waveband \citep[$\sim 20$ for a stochastic model,][]{shao21} \footnote{The delayed model \citep{frye12} and the stochastic model \citep{mand20} are different models determining compact remnant masses and possible natal kicks after the supernova explosions. A detailed description sees also \cite{shao21}.}. Therefore, the detectability of future space-borne GW detectors on BH UCXB-GW sources is not optimistic. However, systematic population synthesis simulations based on our simulated parameter space are required in order to obtain a relatively accurate formation rate and detection number of BH UCXB-GW sources in the Galaxy.

\subsection{Influence of BH masses on the parameter space}
To obtain an initial parameter space (donor-star masses and orbital periods) of progenitors of BH UCXB-GW sources, we take a constant initial BH mass of $8~M_{\odot}$. Based on binary population synthesis simulations, \cite{shao19} found that the mass range of newborn BHs in BH binaries with normal-star companions are $5-16~M_{\odot}$. For a BH with an initial mass of $5~M_{\odot}$, our simulations found that the initial orbital-period range of BH-MS binaries with a $1.2~M_{\odot}$ donor star that can evolve into detectable LISA sources within a distance of 1 kpc are $0.4-2.0$ days, which is slightly wider than that ($0.4-1.83$ days, see also Figure 8) of BH with an initial mass of $8~M_{\odot}$. If BHs possess a high initial mass of $15~M_{\odot}$, the corresponding initial orbital-period range lessen to $0.4-1.6$ days.

Actually, a high BH mass would result in a large orbital angular momentum $J=G^{2/3}M_{\rm bh}M_{\rm d}(M_{\rm bh}+M_{\rm d})^{-1/3}(P_{\rm orb}/2\pi)^{1/3}$. The change rate of the orbital period of a BH binary yielding by angular momentum loss can be written as \citep{chen19}
\begin{equation}
\dot{P}_{\rm orb,am}=3P_{\rm orb}\frac{\dot{J}}{J}.
\end{equation}
For a same orbital period and angular-momentum-loss rate by magnetic braking, a high BH mass naturally results in a large $J$, producing a small $|\dot{P}_{\rm orb,am}|$ ($\dot{P}_{\rm orb,am}<0$). However, the change rate of the orbital period causing by the mass transfer under an assumption of conservative mass transfer is
\begin{equation}
\dot{P}_{\rm orb,mt}=-3P_{\rm orb}\frac{\dot{M}_{\rm d}}{M_{\rm d}}(1-\frac{M_{\rm d}}{M_{\rm bh}}).
\end{equation}
For a same orbital period, donor-star mass, and mass-transfer rate ($\dot{M}_{\rm d}$), a high BH mass tends to produce a large $\dot{P}_{\rm orb,mt}$ ($>0$). As a result of competition between $\dot{P}_{\rm orb,am}$ and $\dot{P}_{\rm orb,mt}$, it would produce system with a wide orbit (Actually, it would produce compact binary system for same initial parameters and low BH mass). To form compact binary system, it requires a small initial orbital period. Therefore, a high initial BH mass tends to reduce initial parameter space of the progenitors of BH UCXB-GW sources.

\section{Conclusion}
Employing the MESA code, we systematically explored the evolution of BH-MS binaries to diagnose whether they can evolve into compact BH-WD binaries, and their descendants can be detected by space-borne GW detectors. Our main conclusions
can be summarized as follows:

\begin{enumerate}
  \item Besides some systems with a Roche-lobe filling donor star at the beginning of binary evolution, other BH-MS binaries with an initial orbital period less than the bifurcation period can evolve into BH UCXBs that can be detected by space-borne GW detectors like LISA. However, only a small fraction of systems with an initial orbital period very close to the bifurcation period can evolve toward low-frequency GW sources whose $\dot{f}$ and chirp masses can be measured.
  \item Through isolated binary evolutionary pathway, BH-MS binaries can not evolve into compact detached BH-WD systems by a stable mass transfer. Our simulations indicate that their descendants are always semi-detached systems, in which a low-mass He core or He WD transfers the material onto the accreting BH.
  \item X-ray luminosities of BH UCXBs detecting by LISA at a distance of 1 kpc are $\sim10^{33}-10^{35}~\rm erg\,s^{-1}$. If the detection horizon is 10 kpc, the luminosities would enhance to be $\sim10^{34}-10^{35}~\rm erg\,s^{-1}$. For an initial orbital period near the bifurcation period, the maximum X-ray luminosity can reach $\sim10^{39}~\rm erg\,s^{-1}$. Certainly, such a high X-ray luminosity requires a fine-tuning $P_{\rm orb,i}$ ($=P_{\rm bif}$) and its luminosity evolution is transient. Therefore, it is mostly likely to detect their electromagnetic counterpart for BH UCXB-GW sources.
  \item The maximum GW frequency of BH UCXBs evolving from BH-MS binaries is about 3 mHz. However, the minimum GW frequency of mass-transferring BH-WD originating from a dynamic process is 6.4 mHz. Therefore, GW frequencies can be used to confirm the evolutionary channel of BH UCXBs.
  \item Within a detection distance of 1 kpc, the initial donor-star masses and initial orbital periods of progenitors of BH UCXBs detecting by LISA are in the range of $0.4-1.6~M_{\odot}$ and $0.2-1.83$ days, respectively. However, the initial parameter space sharply lessen to be $1.0-1.6~M_{\odot}$ and an extremely narrow period range near the bifurcation period for a detection distance of 10 kpc.
  \item Through the isolated binary evolution channel, the birth rate of BH UCXB-GW sources that can be detected by LISA in a distance range from $i-1$ kpc to $i$ kpc is $\sim2.6\times10^{-9}~\rm yr^{-1}$, and the total birth rate is $\sim3.9\times10^{-8}~\rm yr^{-1}$ in the Galaxy. In the future, LISA could detect only a few BH UCXB-GW sources in its mission lifetime.
\end{enumerate}


\acknowledgments {We cordially thank the anonymous referee for deep and constructive comments improving this manuscript. We would
also like to thank Yong Shao for helpful discussions. This work was partly supported by the National Natural Science Foundation of China (under grant Nos. 12273014, and 11733009), and the Natural Science Foundation (under grant number ZR2021MA013) of Shandong Province.}

\end{document}